\documentclass[aps,prb,showpacs,twocolumn,superscriptaddress]{revtex4}

\usepackage[dvips]{graphicx}
\usepackage{epsfig}
\usepackage{dcolumn}
\usepackage{bm}
\usepackage{latexsym}
\usepackage{amssymb}
\usepackage{amsfonts}
\usepackage{amsmath}
\usepackage{color}

\usepackage[colorlinks,bookmarks=false,citecolor=blue,linkcolor=red,urlcolor=blue]{hyperref}
\usepackage{verbatim}

\def\bi{\begin{itemize}}
\def\ei{\end{itemize}}
\def\be{\begin{equation}}
\def\ee{\end{equation}}
\def\bs{\begin{subarray}}
\def\es{\end{subarray}}
\def\ba{\begin{eqnarray}}
\def\ea{\end{eqnarray}}
\def\bd{\begin{displaymath}}
\def\ed{\end{displaymath}}

\def\mc{\mathcal}

\newcommand{\tonda}[1]{ \left( #1 \right) }

\begin{document}

\title{Spin-charge decoupling and the photoemission line-shape 
in one dimensional insulators}
\author{Valeria Lante}\email{valeria.lante@uninsubria.it}
\affiliation{CNISM and Dipartimento di Fisica e Matematica, Universit\`a dell'Insubria, Via Valleggio 11, I-22100 Como, Italy}
\author{Alberto Parola}
\affiliation{CNISM and Dipartimento di Fisica e Matematica, Universit\`a dell'Insubria, Via Valleggio 11, I-22100 Como, Italy}

\date{\today}
         
\begin{abstract}
The recent advances in angle resolved photoemission techniques allowed the 
unambiguous experimental confirmation of spin charge decoupling in quasi
one dimensional (1D) Mott insulators. This opportunity stimulates a quantitative analysis 
of the spectral function $A(k,\omega)$ of prototypical one dimensional correlated models. 
Here we combine Bethe Ansatz results, Lanczos diagonalizations and field
theoretical approaches to obtain $A(k,\omega)$ for the 1D Hubbard model 
as a function of the interaction strength. By introducing a {\it single spinon approximation},
an analytic expression is obtained,
which shows the location of the singularities and allows, when supplemented by 
numerical calculations, to obtain an accurate estimate 
of the spectral weight distribution in the $(k,\omega)$ plane. 
Several experimental puzzles on the observed 
intensities and line-shapes in quasi 1D compounds, like
${\rm SrCuO_2}$, find a natural explanation in this theoretical framework.
\end{abstract}


\maketitle

\section{Introduction}
Since the theoretical prediction of the decoupling of spin and charge excitations
in one dimensional (1D) models \cite{SC}, many experiments have long 
sought to verify this effect \cite{kim}.
According to the spin-charge separation scenario, the vacancy ($e^+$) created by 
removing an electron in a photoemission experiment 
decays into two collective excitations (or {\it quasi-particles}), known as {\it spinon} ($s$) and 
{\it holon} ($h$), carrying spin and charge degrees of freedom respectively.
The recent observation of a well defined two-peak structure in the angle-resolved
photoemission  spectra (ARPES) of the quasi-1D materials SrCuO$_2$ and Sr$_2$CuO$_3$ 
\cite{kim06,Kidd}
is deemed a significant clue of spin-charge decoupling, confirming previous expectations. 

However, other quasi one dimensional materials \cite{other} fail to show 
distinct holon and spinon peaks, 
casting some doubt on the interpretation of ARPES experiments based
on spin charge decoupling. A number of puzzling features also 
suggest that more physics, beyond the simple decay $e^+ \to s + h$, 
is involved in the photoemission process:
the spectral functions of SrCuO$_2$ and Sr$_2$CuO$_3$ reported 
by Kim {\it et al.} \cite{kim06} and by Kidd {\it et al.} \cite{Kidd} 
systematically display broad line-shapes in contrast to the sharp
edges expected on the basis of the available calculations on model systems. The spectral intensity 
also appears considerably weaker in a half of the Brillouin zone, a feature often 
ascribed to cross section effects \cite{suga}. 

A quantitative theoretical understanding of ARPES in low dimensional systems 
is important and deserves a careful investigation because ARPES provides a 
direct experimental probe to the single particle excitation spectrum, allowing 
for reliable estimates of the key parameters governing the physics of strongly 
correlated electrons: the electron bandwidth and the Coulomb repulsion. 
Here we will focus on the  1D Hubbard model, a simple lattice model
defined by just two coupling constants: the nearest neighbor hopping integral $t$ and
the on-site Coulomb repulsion $U$: 
\begin{equation}
H=-t\sum_{i,\sigma} \left [c^\dagger_{i+1,\sigma}c_{i,\sigma} + h.c.\right ] + U\sum_i 
n_{i\uparrow}n_{i\downarrow}
\label{hub}
\end{equation}
Although several other terms,
such as next-nearest hopping, further orbital degrees of freedom, 
temperature, disorder or lattice instabilities,  
would be necessary in a realistic model of these materials,
we believe that an accurate investigation of the simplest hamiltonians should be performed
before facing more challenging problems.

The theoretical studies aimed at the investigation of the 
spectral properties of one dimensional models are either fully numerical, 
like Lanczos diagonalizations \cite{kim} and
density matrix renormalization group (DMRG) techniques \cite{Matsueda},
or are carried out in the limiting cases of infinite \cite{SP_92} 
or vanishing \cite{ET} interaction $U/t$. In the former case, they suffer from severe finite
size effects, in the latter the interplay between charge fluctuations and strong correlations 
is not satisfactorily taken into account. Monte Carlo studies of dynamical properties 
of quantum systems are instead hampered by 
the necessity to perform an analytic continuation to real times. 

In this paper, we provide the first quantitative 
evaluation of the full spectral function $A(k,\omega)$
of the 1D Hubbard model at half filling for intermediate and strong coupling $U/t$ \cite{noteU}.
A formalism based on the Bethe Ansatz solution \cite{LW}, 
and supplemented by Lanczos diagonalizations,
is developed and is shown to provide a transparent description of the dynamical properties of 
mobile charges in Mott insulators.
From this analysis we find that the 1D Hubbard model does indeed contain the physics required
for a quantitative interpretation of photoemission experiments.
In particular:
$i)$ the underlying free electron Fermi surface plays a key role
in defining the shape and the intensity of the ARPES signal, up to fairly large effective couplings 
$U/t$; $ii)$ the
power-law singularities which characterize the spectral function in one dimension give 
rise to intrinsically broad peaks, whose width is proportional to the intensity of the line; $iii)$ 
ARPES data are extremely sensitive to the Hubbard parameters and allow 
for a direct determination of the effective coupling constants in quasi 1D materials. 
As a working example, we apply our method to ${\rm SrCuO_2}$, where accurate 
ARPES data are available \cite{kim}, and we derive reliable estimates for $t$ and $U$. 

The plan of the paper is as follows.
In Section \ref{method_sec} we introduce and motivate the single spinon approximation
which lies at the basis of our method, deriving the predicted formal structure of the
spectral function in one dimensional models.
Section \ref{lanczos_sec} shows how Lanczos diagonalizations provide a precise quantitative
estimate of the quasi-particle weight required for the evaluation of the spectral function.
Then, in Section \ref{weak_sec} we discuss the weak coupling limit, where a thorough field
theoretical analysis is available. The application to the case of ${\rm SrCuO_2}$ is performed
in Section \ref{result_sec}, while in the Conclusions we briefly discuss the  
generalization of our method to more complex one dimensional hamiltonians.

\section{The analytical structure of the spectral function}
\label{method_sec}
The dynamical properties of one (spin down) hole
in the half filled Hubbard model are embodied in the spectral function $A(k,\omega)$
which, at zero temperature, can be written as:
\be 
A(k,\omega)=\sum_{\{\vert\Psi_1\rangle\}}\vert\langle 
\Psi_1\vert c_{k,\downarrow}\vert\Psi_0\rangle\vert^2\delta\tonda{\omega-E_1+E_0}
\label{ako}
\ee
where $\vert\Psi_0\rangle$ is the ground state of the model at zero doping, i.e. when  
the number of electrons $N$ equals the number of sites of the lattice $L$, 
$E_0$ is the corresponding energy and 
$\{\vert\Psi_1\rangle \}$ represents a complete set of one-hole intermediate
states, of energies $E_1$. 
The whole energy spectrum of the Hubbard hamiltonian (\ref{hub})
can be obtained from the Lieb and Wu equations \cite{LW}. In the thermodynamic
limit, its structure has been 
thoroughly investigated in a series of papers by Woynarovich \cite{woy2,woy} 
(see also the comprehensive book by Essler {\it et al.} Ref. \cite{essler}). 
In summary, the exact excitation spectrum at half filling 
and at arbitrary coupling $U/t$ depends on two sets of ``rapidities" 
describing the charge and spin degrees of freedom respectively. The excitation energy is
always written as the sum of contributions involving 
just two elementary excitations, representing collective quasi-particles: 
``holons" (of momentum $k_h$ and energy $\epsilon_h(k_h)$)
and ``spinons"(of momentum $Q \in (\frac{\pi}{2},\frac{3\pi}{2})$ 
and energy $\epsilon_s(Q)$). The simplest physical 
excitation created by the removal of an electron of momentum $k$ 
gives rise to one holon and one spinon satisfying the momentum conservation equation $k=k_h+Q$.
The total energy of this state is $E_1=E_0+\epsilon_h(k_h)+\epsilon_s(Q)$. 
Besides this suggestive ``decay" mechanism of the electron, other excited states 
also appear in the exact spectrum: they are either multi-spinon and multi-holon states,
or states involving the creation of double occupancies \cite{woy}.
However, it is remarkable that the full excitation spectrum 
can be always expressed in terms of $\epsilon_h(k_h)$ and $\epsilon_s(Q)$, showing that 
spin charge decoupling holds, in the Hubbard model, 
at all values of $U/t$ and at all energy scales \cite{essler}.
The two quasi-particles,
holon and spinon, are both collective excitations involving an extensive number of degrees
of freedom and can be approximately related to simple real space pictures of a 
``hole" and an unpaired spin only in the strong coupling limit, where spin-charge
decoupling acquires a more intuitive meaning. 
As $U\to 0$ the holon and spinon bands reduce to simple analytical forms \cite{LW}, closely
related to the free particle band structure:
$\epsilon_h(k_h)=4t\,\cos(k_h/2)$ and $\epsilon_s(Q)=2t\,\vert \cos Q \vert $.

While the whole energy spectrum of the Hubbard hamiltonian is known in detail,
the matrix elements appearing in Eq. (\ref{ako}) are of difficult evaluation. Moreover
the summation over the intermediate states formally 
involves a number of terms exponentially large in $N$, making the exact implementation of
the definition (\ref{ako}) impractical.
Our approach, which allows for the evaluation of the full spectral function in 
the thermodynamic limit,
is based on the {\it single spinon approximation}: i.e. we neglect the contribution 
to the spectral function coming from all multi-spinon 
excited states and all excitations with complex rapidities, but we evaluate exactly the
matrix elements involving one holon and one spinon. The accuracy of this method is
tested {\it a posteriori} by use of a completeness sum rule and can be estimated 
of the order of few percents. Such a 
remarkable performance of the single spinon approximations is not unusual in one dimensional
physics: a known example is provided by the Haldane-Shastry spin model (HSM) \cite{haldane}, 
where each intermediate state contributing to the dynamical spin correlation is
completely expressible in terms of eigenstates of the HSM with only two spinons. 
In this case, only a small $O(L)$ number of eigenstates contribute
to the {\it exact} dynamical spin correlation function as proved in Ref. \cite{haldanez}.
Similarly, in our approach, the most relevant intermediate states are 
expressible is terms of eigenstates of the Hubbard model with only one spinon
and one holon excitations.

A first clue on the structure of the spectral function in one dimensional models can be obtained
by analyzing the $U\to\infty$ limit, where double occupancies are inhibited and 
several exact results are available \cite{SP_92}.
At half filling ($N=L$) the Hubbard hamiltonian is mapped
onto a Heisenberg hamiltonian: each site is singly occupied and
the ground state is a non-degenerate singlet of zero momentum \cite{note}.
When a hole of momentum $k_h$ is created, all the eigenfunctions of the Hubbard 
hamiltonian (with periodic boundary conditions) can be written as \cite{Ogata}:
\be\label{s}
\vert \Psi_{1} >=\frac{1}{\sqrt L}\sum_{x,\{y_i\}} 
e^{ik_hx}\phi_H(y_1,\ldots, y_M)\,\vert x, \{y_i\} \rangle
\ee
where $\vert x, \{y_i\} \rangle$ represents the configuration of $L-1$ electrons
defined by the positions of the $M=L/2$ spin up ($\{y_i\}$) and of the hole ($x$). The amplitude 
$\phi_H$ is a generic eigenfunction of the Heisenberg hamiltonian on the ``squeezed chain", 
i.e. on the $L-1$ site ring defined by all the sites occupied by an electron.
The intermediate states $\vert \Psi_1>$ entering the spectral function (\ref{ako})
have momentum $-k$ relative to the ground state at half filling. Due to the factorized form 
of the eigenfunctions (\ref{s}) the total momentum of the state satisfies
$k=k_h+Q$ where $Q$ is the momentum of the Heisenberg eigenfunction $\phi_H$, expressed in 
integer multiples of $2\pi/(L-1)$ in finite chains. In the thermodynamic limit 
the energy of the intermediate state is $E_1=E_0+\epsilon_h(k_h)+\epsilon_{s}(Q)$,
where, to lowest order in $J=4\,t^2/U$, the first (holon) contribution is 
just the kinetic energy of a free particle ($\epsilon_h(k_h)=2t\cos{k_h}$)
and the second one (spinon) is the energy of the eigenstate $\phi_H$ 
referred to the ground state energy of the Heisenberg ring of $L$ sites \cite{spinon}. 
This analysis shows, in an intuitive way, the origin of momentum and energy conservation
in the decay process of the vacancy and suggests that, in the $U\to\infty$ limit, the most relevant 
contributions to the sum of intermediate states in Eq. (\ref{ako}) come from the lowest energy 
eigenstates $\phi_H$ of the Heisenberg hamiltonian for the $L-1$ allowed momenta 
$Q=\frac{2\pi}{L-1}n$ ($n=0\dots L-2$). Accordingly, the sum over an exponentially large set of 
eigenstates $\{\vert \Psi_1 >\}$ in Eq. (\ref{ako}) can be (approximately) 
replaced by a sum over $L-1$ {\it single spinon states}. 
This special set of intermediate states $\vert \Psi_1\rangle$, 
which we argue provides the dominant contribution to 
the spectral weight for each spinon momentum $Q$, will be referred to as 
$\vert -k,Q\rangle$ in order to emphasize the
two quantum numbers which uniquely identify them. The 
single spinon approximation can be easily tested in the $U\to\infty$ limit \cite{SP_92}
where it proves extremely accurate.
In the next Section we will show that it remains fully satisfactory also at finite 
coupling. 
In fact, it is known \cite{essler} that the eigenstate structure 
of the Hubbard model displays a remarkable continuity in $U/t$, the only singular 
point being the (trivial) free particle limit $U=0$. 
However,  
when charge fluctuations are allowed for, by lowering the strength of the on-site 
repulsion $U$, the identification of the single spinon states $\vert -k,Q\rangle$ 
is not easy, because the spinon momentum $Q$ is not a good quantum number any more, 
although it can be still formally defined on the basis of the Bethe Ansatz solution of the
Hubbard model \cite{woy}. The key observation, which will be exploited 
in the next Section, is that the correct single spinon states can be identified
at finite $U$ via Lanczos or DMRG calculations by following {\it adiabatically} the evolution 
of the Heisenberg states as $U$ is gradually decreased. 

\begin{figure}
\includegraphics[width=0.475\textwidth]{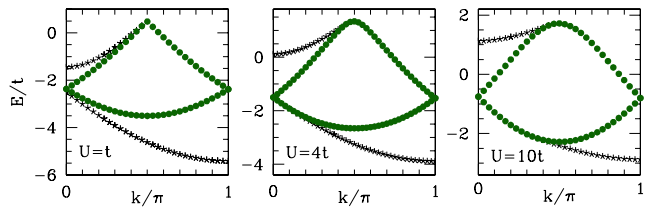}
\caption{\label{singularities}Location of the singularities of the spectral function in the
thermodynamic limit for three values of $U$ in the plane ($k$,$E=-\omega$).
The (green) circles correspond to the singularities
of the $\mc Z_k(Q)$ ({\it holon branch}), while the black stars
to the extrema of the excitation spectrum.}
\end{figure}

Keeping only the single spinon states in the summation of Eq. (\ref{ako}),
the spectral function in the thermodynamic limit becomes
\begin{eqnarray}
A(k,\omega) &=&
\int \frac{dQ}{2\pi} \,{\mc Z}_k(Q) \, 
\delta \left({\textstyle \omega-\epsilon_h(k-Q)-\epsilon_{s}(Q)}\right ) 
\nonumber\\
&=&\frac{1}{2\pi}\,\sum_{Q^*}\frac{{\mc Z}_k(Q^*)}{\vert v_h(Q^*-k)+v_s(Q^*)\vert}
\label{a3}
\end{eqnarray}
Where we have defined the {\it quasi-particle} weight as the matrix element
\be\label{Z}
\mc{Z}_k(Q)\equiv \lim_{L\to\infty} 
(L-1)\,\vert \langle -k,Q\vert c_{k,\downarrow} \vert\Psi_0 \rangle\vert^2
\ee
The sum in Eq. (\ref{a3}) runs over all the solutions $Q^*(k,\omega)$ of the algebraic equation
\be
\omega=\epsilon_h(k-Q)+\epsilon_{s}(Q)
\label{omega}
\ee
where $\epsilon_h(k_h)$ [$\epsilon_{s}(Q)$] is
the known holon [spinon] excitation energy \cite{woy} and
$v_h(k_h)=\frac{d\epsilon_h}{dk_h}$ [$v_s(Q)=\frac{d\epsilon_s}{dQ}$] the associated velocity.
Equation (\ref{a3}) is the main result of this work: an explicit and computable
expression for the spectral function of one dimensional models. In the 
special case of the Hubbard model, the Bethe Ansatz
solution directly provides spinon and holon dispersions in the thermodynamic limit
further simplifying the evaluation of the spectral function.
Due to the presence of a spinon Fermi surface, 
the dispersion relation $\epsilon_s(Q)$ is defined 
only in the interval $\frac{\pi}{2} < Q < \frac{3\pi}{2}$, \cite{noteQ}
it vanishes at the boundaries and has a single maximum at $Q=\pi$, 
while $\epsilon_h(k_h)$ is an even and periodic function in the whole range 
$-\pi < k_h < \pi$ with maximum at $k=0$ \cite{woy2,woy}. 
The only missing ingredient in Eq. (\ref{a3}) is the quasi-particle weight $\mc Z_k(Q)$ which 
defines the line-shape and intensity of the spectral function. Previous studies \cite{SP_98}
have shown that
in spin isotropic models, like the Hubbard model, the quasi-particle weight is a regular function
with square root singularities at the spinon Fermi surface $Q=\pi\pm \pi/2$. This implies that 
$A(k,\omega)$ has power law singularities too, whenever either $Q^*$ defined by Eq. (\ref{omega})
lies at the spinon Fermi surface, or when the total excitation velocity $v_h(Q^*-k)+v_s(Q^*)$
vanishes. In both instances, square root divergences are expected \cite{note2}: 
in the former case the location of the singularity identifies the holon 
dispersion via (\ref{omega}) $\omega=\epsilon_h(k+\pi\pm\frac{\pi}{2})$; in the latter case
the singularity is trivially due to band structure effects and does not necessarily 
corresponds to a pure spinon contribution as often assumed. However, at small to moderate 
interactions $U/t$, the holon velocity $|v_h(k_h)|$ displays an abrupt drop around $k_h\sim \pi$
\cite{woy} placing the band lower edge close to $Q\sim \pi+k$, i.e. at 
$\omega\sim \epsilon_h(\pi) + \epsilon_s(\pi+k)$, thereby following the spinon band
for $ 0 < k < \frac{\pi}{2}$. This particular feature of the Hubbard model dispersion 
is apparent in the shape of the holon spectrum \cite{woy2} which sharply bends at $k_h\sim \pm\pi$
so to display a vanishing charge velocity at band edges. This also agrees with the ``relativistic"
form of the holon spectrum predicted by bosonization at weak coupling \cite{ET}, 
as reported in Eq. (\ref{lorentz}).
The expected location of the square root singularities of the
spectral function in the $(k,\omega)$ plane is shown for few values of the coupling in
Fig. \ref{singularities}. The holon branch (shown as full circles in the figure) 
marks precisely the holon excitation spectrum $\epsilon_h(k_h)$ while the 
location of the singularities due to the band structure (shown as crosses in the figure)
differs from the spinon $\epsilon_s(Q)$ dispersion by less than $0.1 t$. 
Note also that 
the curvature of the ``spinon branch" displays a significant dependence on $U/t$, allowing for 
a rather precise experimental determination of the effective coupling ratio.
Therefore we conclude that precise photoemission data, able to identify the 
singularities of the spectral function, do provide direct information on both holon
and, within a good approximation, also spinon excitations. 

\begin{figure}
\includegraphics[width=0.475\textwidth]{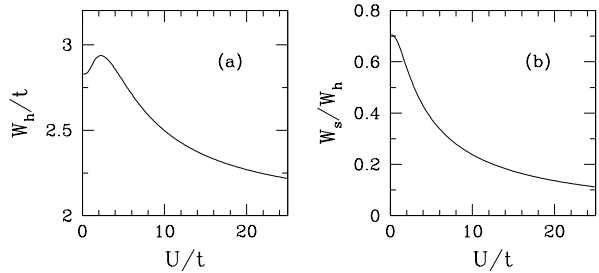}
\caption{\label{bandwidth}Panel (a): holon bandwidth $W_h=\epsilon_h(\pi/2)-\epsilon_h(\pi)$
as a function of $U/t$. Panel (b): ratio between the spinon bandwidth $W_s=\epsilon_s(\pi)$ 
and $W_h$ as a function of $U/t$.
}
\end{figure}

The full holon bandwidth is always $4t$ at all couplings, due to the
particle-hole symmetry of the Hubbard model but the upper and lower branches
of the holon band are not symmetrical at finite $U$. This observation is relevant for the
correct interpretation of photoemission experiments, because an estimate of the 
effective hopping integral $t$ is usually performed by measuring the {\it half bandwidth} of the 
upper holon branch \cite{note3} leading to a sizable overestimate of $t$. 
In Fig. \ref{bandwidth} we show the bandwidth $W_h$ of the upper 
holon branch (i.e. $\epsilon_h(\pi/2)- \epsilon_h(\pi)$) and the ratio between 
the spinon and the holon bandwidths $W_s/W_h$ as a function of the coupling $U/t$. 
Both quantities, which allow for a direct estimate of
$t$ and $U/t$ from ARPES, show a remarkable (even non monotonic) 
dependence on the coupling constants. 
By comparing these results with the dispersion curves for SrCuO$_2$ reported in Ref. \cite{kim06}
we can estimate the Hubbard effective coupling constants 
appropriate of this material: $t\sim0.53$ eV and $U/t\sim 7$.

\section{The quasi-particle weight from Lanczos diagonalization}
\label{lanczos_sec}
Unfortunately, the formal Bethe Ansatz solution does not lead to a practical way for the 
evaluation of the quasi-particle weight (\ref{Z}) at arbitrary couplings 
and therefore we resort to Lanczos diagonalizations
in lattices up to $L=14$ sites. As previously noticed, we first have to devise 
a method to select the correct single spinon states at finite coupling $U/t$, for these
states are not identified by a good quantum number at finite $U$. Our method is based on an 
adiabatic procedure starting from the strong coupling limit.
We first perform a Lanczos diagonalization on the $(L-1)$-site
Heisenberg model in the symmetry subspace of total momentum $Q$, leading to the numerical
determination of $\phi_H$ and of the exact eigenstates of 
the one-hole Hubbard model for $U\to\infty$ via Eq. (\ref{s}).
In this limit, the single spinon states are indeed 
the lowest energy eigenstates at fixed spinon momentum $Q$ and can be easily obtained by Lanczos 
(or DMRG) technique, whilst at finite $U$ the relevant intermediate states are not necessarily 
in the low excitation energy portion of the Hubbard spectrum. 
Then we take advantage of the
continuity of the one spinon states between the weak and strong coupling limit
by adiabatically lowering the interaction strength $U$ and 
performing successive Lanczos diagonalizations for smaller and smaller couplings $U_n$.
At the $n^{th}$ step we keep the exact eigenstate having the largest overlap with the 
eigenstate at the $(n-1)^{th}$ level. In this way we are able to identify the
single spinon states down to small values of $U\sim t$, each state being uniquely identified 
by $Q$, i.e. by the momentum of the ``parent" Heisenberg eigenstate. 

A check on the validity of the single-spinon approximation comes from the
completeness condition on the intermediate states:
\begin{eqnarray}
n_\downarrow(k) &=& \langle \Psi_0\vert c_{k,\downarrow}^{\dagger} c_{k,\downarrow} \vert\Psi_0\rangle
= \sum_{\{\vert\Psi_1\rangle\}}\vert\langle \Psi_1\vert c_{k,\downarrow}\vert\Psi_0\rangle\vert^2
\nonumber \\
&\ge& \sum_Q \vert \langle -k,Q \vert  c_{k,\downarrow} \vert \Psi_0 \rangle \vert^2 
\nonumber \\
&=& 
\frac{1}{L-1}\sum_Q \mc Z_k(Q)
\label{completeness}
\end{eqnarray}
where $n_\downarrow(k)$ is the momentum distribution of the down spins
at half filling
and the equality holds if and only if the single spinon states included in the sum 
via the definition of the quasi-particle weight $\mc Z_k(Q)$ (\ref{Z}) 
exhaust the spectral weight at each $k$. The amount of violation of this
sum rule quantifies the weight of {\it all} the neglected states in the Hilbert space
due to the single-spinon approximation.
\begin{figure}
\includegraphics[width=0.4\textwidth]{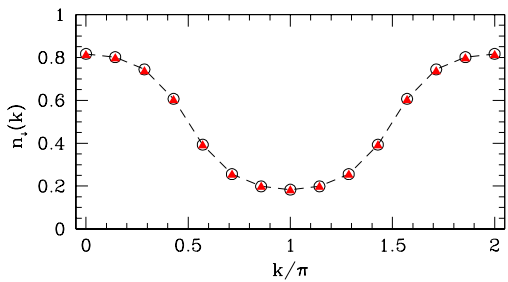}
\caption{\label{momentum distribution}Momentum distribution $n_\downarrow(k)$
(black open circles) and $\frac{1}{L-1}\sum_Q \mc Z_k(Q)$ (red triangles) 
versus the hole momentum $k$ from Lanczos diagonalization, for $U=7t$ in a $L=14$ ring. 
}
\end{figure}
In Fig.\ref{momentum distribution} we plot $n_\downarrow(k)$ and
$\frac{1}{L-1}\sum_Q \mc Z_k(Q)$ restricted to the one spinon states: the
violation of the completeness condition is smaller than $0.01$ at all $k$'s \cite{notek}.
Note how, even at fairly large values of $U/t$, the momentum distribution 
is considerably depressed for $k$ larger than 
the free electron Fermi momentum $k_F=\frac{\pi}{2}$, strongly 
reducing the spectral weight in the second half of the Brillouin zone. This feature
is consistent with the photoemission experiments performed with high energy photons 
\cite{kim06,suga}. Conversely, 
in the strong coupling limit $U\to \infty$, the momentum distribution becomes flat,
$n_\downarrow(k)=1/2$, washing out this effect. 

\begin{figure}
\includegraphics[width=0.475\textwidth]{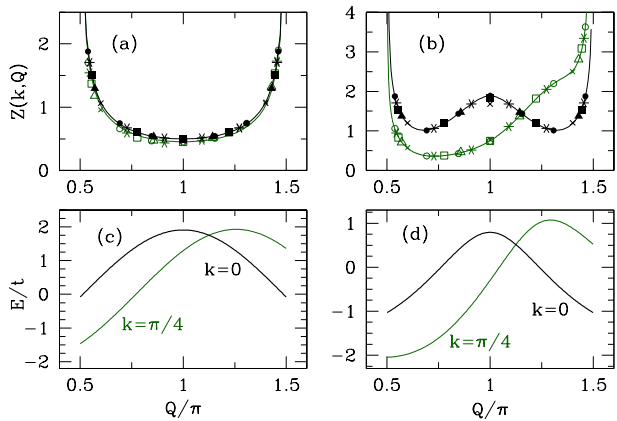}
\caption{\label{z0}
Panel (a):
$\mc Z_k(Q)$ versus spinon momentum $Q$ for $U=100t$ and
different lattice sizes ($\times$: L=6, $\blacktriangle$: L=8,
{\small $\blacksquare$}: L=10, $\ast$: L=12, $\bullet$: L=14).
Open (green) symbols for total momentum $k=\pi/4$ and full (black) symbols for $k=0$.
Lines are polynomial fit to Lanczos data. 
Skewed boundary conditions are used in order to fix the same total momentum of
the state $-k$ (relative to half filling) for all $L$'s.
Panel (b): same as (a) for $U=7t$.
Panel (c): binding energy at fixed $k$ referred to half filling versus spinon
momentum $Q$ in the thermodynamic limit $E=E_0-E_1$ for $U=100t$.
Panel (d): same as (c) for $U=7t$.
}
\end{figure}

The dependence of the quasi-particle weight on the strength of the
Coulomb repulsion has been investigated and is summarized in Fig.\ref{z0}
for strong and intermediate $U/t$ and for different lattice sizes.
The quasi-particle weight has been evaluated by Lanczos diagonalization
on lattices ranging from $L=6$ to $L=14$ sites. By using 
standard periodic boundary conditions, the total momentum of the state 
would be quantized in units of $2\pi/L$, making size scaling 
impractical. In order to avoid this problem we have adopted 
skewed boundary conditions: given an arbitrary hole momentum $k$ we choose 
the flux at the boundary in such a way to match $k$ with the quantization rule. 
Figure \ref{z0} reveals an astoundingly negligible size dependence and the expected 
vanishing of the quasi-particle spectral weight outside the spinon Fermi surface,
with singularities at the Fermi momenta.
While $\mc Z_k(Q)$ is almost independent on $k$ at large $U$,
as expected \cite{SP_92}, it shows more structure for realistic values of $U/t$.
The further peak (or shoulder) present for $k\le \frac{\pi}{2}$ is indeed 
reminiscent of the free Fermi nature of the electrons at $U=0$. 
In the free particle {\it limit}, only one state provides a finite contribution
to the spectral function: the holon sits at the bottom of the band ($k_h=\pi$)
and the quasi-particle weight $\mc Z_k(Q)$ reduces to a delta function
at $Q=\pi+k$. When such a form of $\mc Z_k(Q)$ is substituted in Eq. (\ref{a3}), the
known free particle result is recovered.
Remarkably, a remnant of the free particle peak in $\mc Z_k(Q)$
is still visible at $U=7t$, as shown in Fig. \ref{z0}b.

\section{Weak coupling limit}
\label{weak_sec}
\begin{figure}
\includegraphics[width=0.475\textwidth]{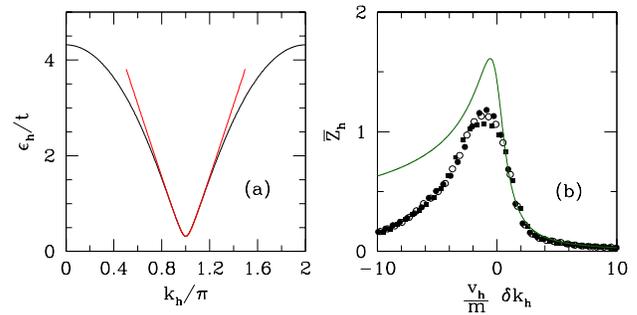}
\caption{\label{lorentz_fig}
Panel (a): Holon spectrum $\epsilon_h(k_h)$ of the Hubbard model at $U=3t$ from Bethe Ansatz 
(shifted by $\mu=U/2$) compared to the Lorentz form (\ref{lorentz}). The fitting parameters are 
$m=0.316 t$ and $v_h=2.43 t$.
Panel (b): Dimensionless holon quasi-particle weight $\bar Z_h = \sqrt{\frac{m}{v_h}}Z_h$ 
in the single spinon approximation for $U=3t$ and
lattice sizes $L=10$ (full squares), $L=12$ (empty circles), $L=14$ (full circles), compared to
the bosonization result (\ref{zh}) (line). 
}
\end{figure}

The Green's function of one dimensional models has been thoroughly investigated by bosonization
methods: while in the Luttinger liquid regime its asymptotic form is characterized by power law 
tails \cite{SC}, precisely at half filling the Green's function is known to display a more 
complex behavior due to the presence of a gap in the holon spectrum. At weak coupling,
the holon dispersion near the bottom of the band shows a ``relativistic " structure:
\be
\epsilon_h(k_h) = \sqrt{v_h^2\delta k_h^2+m^2} 
\label{lorentz}
\ee  
where $m$ is the charge gap and $\delta k_h=k_h\pm\pi$ 
is the holon momentum measured from the bottom of the band. 
Note that the holon spectrum (\ref{lorentz}) is shifted by $\mu=U/2$
with respect to our previous definition. In Fig. \ref{lorentz_fig}a we 
plot the exact Bethe Ansatz spectrum at $U=3t$ and the form (\ref{lorentz}) 
predicted by bosonization with suitably chosen parameters $m$,$v_h$. 

In order to compare the results of our single spinon approximation with the bosonization
form, it is convenient to introduce the single hole Green's function in imaginary time:
\be
G_\downarrow(k,\tau) = <\Psi_0 \vert c^\dagger_{k,\downarrow}\,e^{-(H-\mu)\tau}\,c_{k,\downarrow}
\vert \Psi_0>
\label{green}
\ee
According to bosonization \cite{ET}, the Green's function $G_\downarrow(k,\tau)$ of a 
hole of momentum close to $k_F=\pi/2$ acquires a factorized form in real space: 
\begin{eqnarray}
G^R_\downarrow(x,\tau) &\equiv& \int \frac{dk}{2\pi} G^R_\downarrow(k,\tau) 
e^{ikx} \nonumber \\
&=& e^{i\frac{\pi}{2}x}\,G_h(x,\tau)\,G_s(x,\tau)
\label{factor}
\end{eqnarray}
where the superscript $R$ identifies the contribution to the Green's function due to {\it right moving} 
holes. Here, $G_h$ and $G_s$ just depend on holon and spinon degrees of freedom respectively.
The spinon term is simply given by
\be
G_s(x,\tau) = \frac{1}{\sqrt{v_s\tau+ix}}
\label{gs}
\ee 
while the holon contribution is predicted, by the form factor approach, to behave as
\be
G_h(x,\tau) = \Gamma\,\sqrt{\frac{m}{v_h}}\,\int_{-\infty}^\infty \,d\theta \, 
e^{\left [\theta/2 -m\tau\cosh\theta -im\frac{x}{v_h}\sinh\theta\right ]}
\label{gh}
\ee 
with $\Gamma\sim 0.0585...$ \cite{ET}. 
The question now arises whether our single spinon approximation is 
consistent with such a factorized form.
By inserting a complete set of intermediate states into the definition (\ref{green}) 
and adopting the single spinon 
approximation, the full Green's function in imaginary time can be written as
\be
G_\downarrow(k,\tau) = \frac{1}{L}\sum_Q {\mc Z_k(Q)} 
e^{-\left [\epsilon_h(k_h)+\epsilon_s(Q)\right ]\tau}
\label{green2}
\ee
where the momentum conservation relation $k=k_h+Q$ is understood and the holon spectrum
$\epsilon_h(k_h)$ is now referred to the chemical potential $\mu$. Notice that,
due to momentum conservation, the combined requirements of having $k\sim k_F=\frac{\pi}{2}$ and
$k_h\sim -\pi$ (i.e. the hole sits near the bottom of the band)
force $Q\sim \frac{3\pi}{2}$. By substituting the asymptotic
forms (\ref{lorentz}) and $\epsilon_s\left ( \frac{3\pi}{2}-q\right)=v_sq$ for $q\gtrsim 0$ we get:
\begin{eqnarray}
\label{ssa}
&&G^R_\downarrow(x,\tau) = \,e^{i\frac{\pi}{2}x}\,\frac{m}{v_h}\,
\int_0^{\pi} \frac{dq}{2\pi} e^{-\left [iqx+v_sq\tau\right ]} \qquad\\
&&\qquad\times\,
\int_{-\infty}^\infty \frac{d\theta}{2\pi}
{\mc Z_k(Q)}
\,\cosh\theta\,e^{\left [ -im\frac{x}{v_h}\sinh\theta - m\tau\cosh\theta\right ]}\nonumber 
\end{eqnarray}
where we set $\delta k_h\equiv -\frac{m}{v_h} \sinh\theta$. This form does indeed factorize in 
a holon and spinon part, as predicted by bosonization, provided the quasi-particle weight does: 
\be
\mc Z_k(Q)\sim Z_h(k_h)\, Z_s(Q)
\label{zfact}
\ee
Notice that our approach, being based on a numerical evaluation of the quasi-particle
weight, does not allow for an independent demonstration of such a factorized form. We just
observe that the bosonization approach and the single-spinon approximation lead to the
same result {\it if we assume} that Eq. (\ref{zfact}) holds. Following Ref. \cite{essler}
we argue that the factorization of the quasi-particle weight at low energies (\ref{zfact}) 
reflects the trivial structure of the holon-spinon scattering matrix in this limit. 

As previously noticed, the spinon contribution to the quasi-particle weight gives rise to the
square root divergence at the spinon Fermi surface, with leading behavior 
$Z_s\left ( \frac{3\pi}{2} -q\right )\sim q^{-1/2}$ for $q\gtrsim 0$ which correctly 
reproduces Eq. (\ref{gs}) when the ultraviolet cut-off in Eq. (\ref{ssa}) is disregarded. 
Matching Eqs. (\ref{gh}) and (\ref{ssa}) then selects a 
unique form of the holon quasi-particle weight:
\be
Z_h(k_h) = \sqrt{2v_h}\,\frac{\sqrt{\epsilon_h(k_h)-v_h\delta k_h}}{\epsilon_h(k_h)}
\label{zh}
\ee
with $\epsilon_h(k_h)$ given by Eq. (\ref{lorentz}). 
The scale factor in Eq. (\ref{zh}) has been fixed by evaluating the Green's function in the
$\tau\to 0$ limit, where it coincides with the momentum distribution.
In Fig. \ref{lorentz_fig}b we compare 
Eq. (\ref{zh}) with the numerical results for $Z_h(k_h)$ obtained by Lanczos diagonalization
at $U=3t$. No fitting parameters have been used:
In order to obtain $Z_h(k_h)$ we first evaluated $\mc Z_k(Q)$, as discussed in 
Section \ref{lanczos_sec}, then we divided the result by $Z_s(Q)\sim q^{-1/2}$ evaluated at the
spinon momentum $Q$ closest to the spinon Fermi point $Q_F=\frac{3\pi}{2}$. The two parameters
$v_h$ and $m$ are independently obtained from the 
holon spectrum (also shown in Fig. \ref{lorentz_fig}a).
As usual the Lanczos data display a very small size dependence and allow for
a precise identification of the holon quasi-particle weight $Z_h(k_h)$. 
The agreement
between the two expressions is remarkable for $\delta k_h >0$ while some discrepancy is found
for negative $\delta k_h$. Note however that the asymptotic form of the
holon quasi-particle weight (\ref{zh}) holds only at low energies and weak coupling,
while the comparison shown in Fig. \ref{lorentz_fig} is performed for $U=3t$. 
The results at lower values of $U/t$ are plagued by severe finite size effects: 
in the $U\to 0$ limit, the holon mass $m$ vanishes exponentially and the dimensionless 
momentum scale $m/v_h$ vanishes as well. Therefore, at very weak coupling, the
relevant holon momenta are constrained in an extremely small interval around $k_h=\pm \pi$,
a range not easily accessible due to the momentum quantization rule in finite Hubbard rings. 

\section{Results for ${\bf SrCuO_2}$}
\label{result_sec}
We are now ready to compare our results for the spectral function of the 1D Hubbard model
with precise photoemission data recently obtained for ${\rm SrCuO_2}$ \cite{kim06}.
A preliminary study, based on the strong coupling limit of the Hubbard model, pointed out
some discrepancies, related to the peak heights and widths \cite{kim06}. 
\begin{figure}
\includegraphics[width=0.475\textwidth]{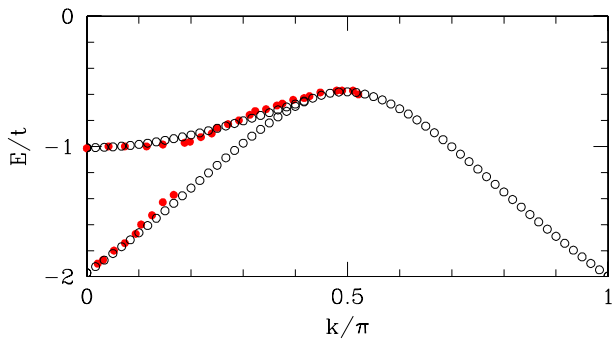}
\caption{\label{kim_fig}
Locus of singularities of the spectral function for the Hubbard model at $U=3.7$ eV and 
$t=0.53$ eV (open circles) compared to the experimental results by Kim {\it et al.} \cite{kim06}
(full circles) for $k_\perp=0.1$.}
\end{figure}
Fig. \ref{kim_fig} shows the singularity loci of the 1D Hubbard model with the parameters
$t=0.53$ eV and $U=3.7$ eV, together with the ARPES results from Kim {\it et al.} \cite{kim06}.
The nice agreement suggests that this material indeed represents a good experimental realization
of the simple one dimensional Hubbard model. The effects due to inter-chain coupling, phonons,
finite temperature and other perturbations appears rather small and mostly limited to the spinon 
branch. We remark that the same material has been already theoretically investigated 
on the basis of the Hubbard and t-J model by several groups \cite{kim06,popovich,koitzsch} 
leading to different sets of parameters both for the hopping integral $0.3 {\rm eV}\lesssim t \lesssim 0.7 {\rm eV}$
and for the Coulomb repulsion $2{\rm eV} \lesssim U \lesssim 6.5{\rm eV}$. 
Our analysis shows that both spin and charge fluctuations play a key role in determining the 
line shape of the spectral function of the Hubbard model, even at moderately high values of the
coupling $U/t$. 
\begin{figure} 
\includegraphics[width=0.475\textwidth]{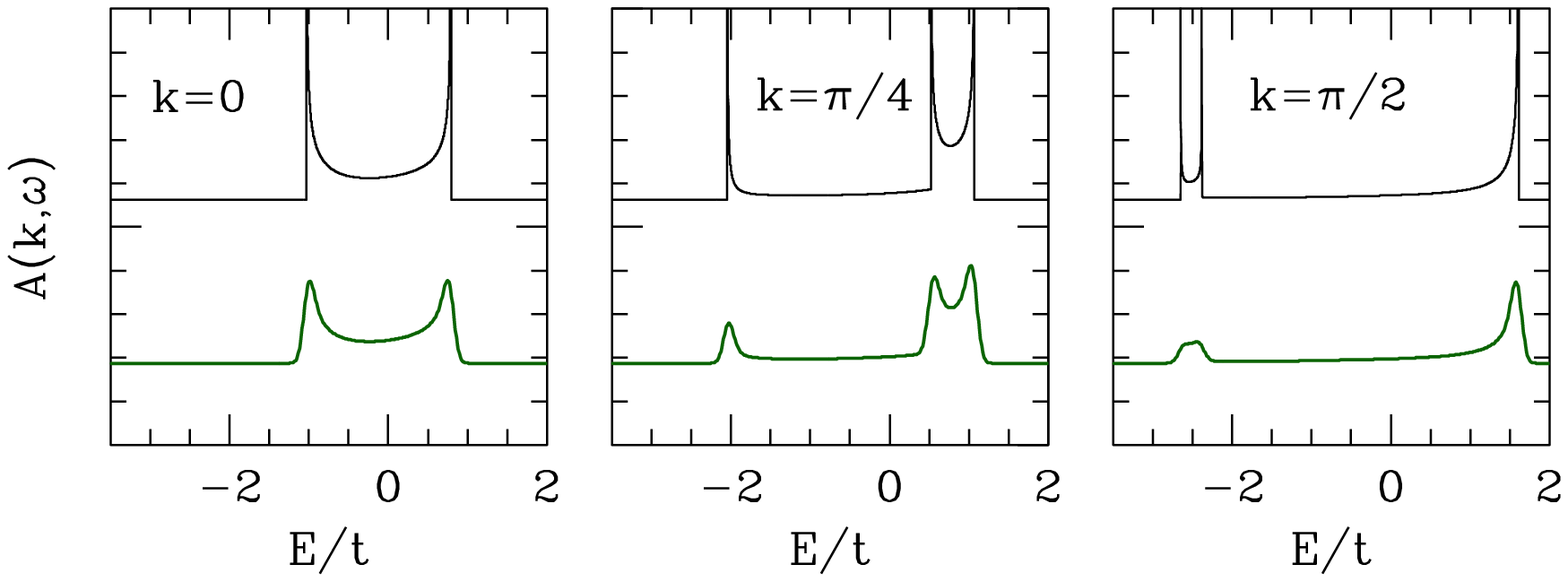}
\includegraphics[width=0.500\textwidth]{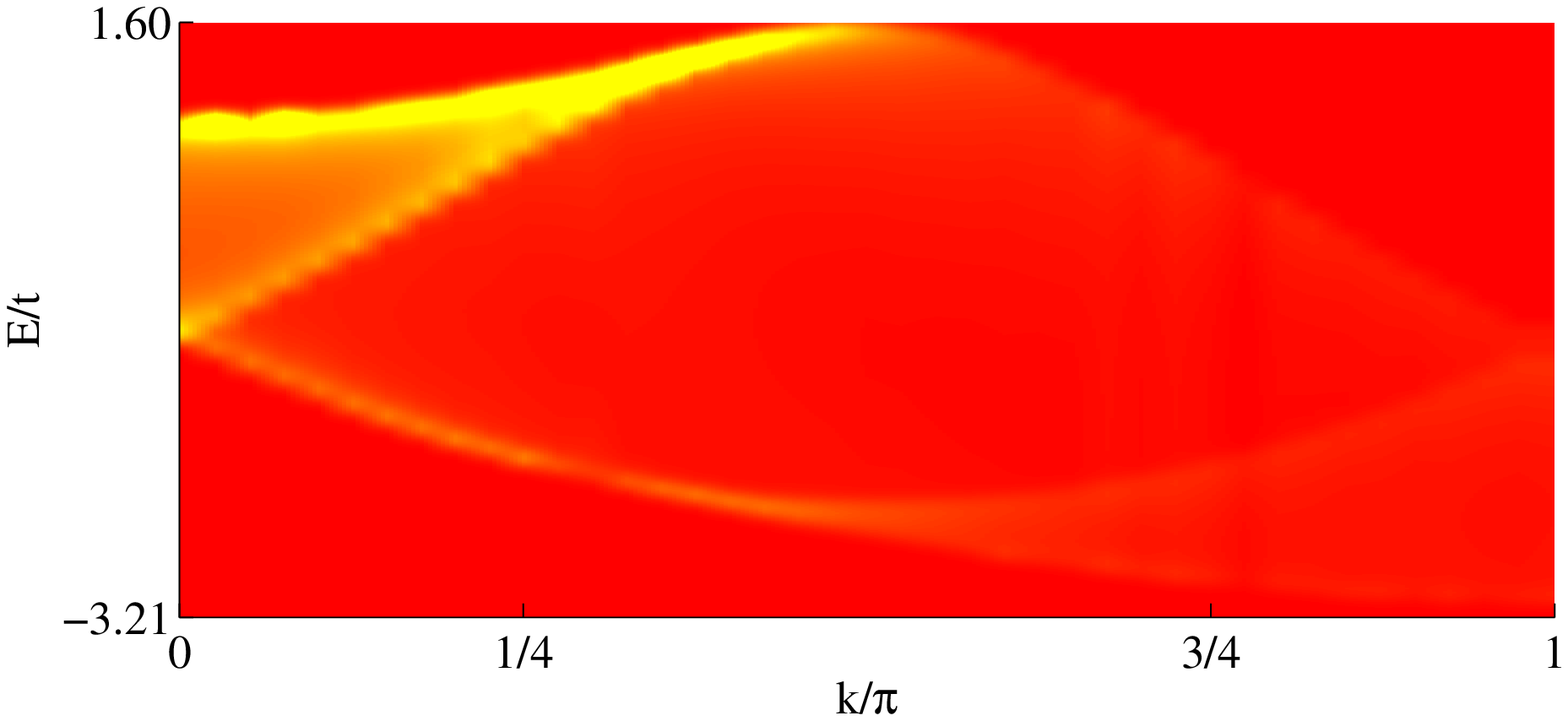}
\caption{\label{spectral_summary.eps} Upper panels: $A(k,\omega)$ calculated
via Eq. (\ref{a3}) for three representative values of the momentum $k$ and $U=7t$.
The binding energy is $E=-\omega$.
The lower curves are the convolution of the spectral function
with a Gaussian with FWHM equal to $0.12t$
corresponding to an experimental resolution of $60$ meV \cite{kim06}.
Lower panel: density plot of the spectral function on the $(k,E)$ plane with 
the experimental broadening.}
\end{figure}

In Fig.\ref{spectral_summary.eps} the spectral function has been plotted
versus the binding energy $E=-\omega$ for three representative values of the
total hole momentum $k$. The experimental line broadening reported in Ref.\cite{kim06}
has been also included in the Hubbard model results, leading to a merging of close peaks.
The density plot clearly reproduces the overall
shape defined by the singularities of the spectral function shown in Fig.\ref{kim_fig}. 
As expected, most of the spectral weight is indeed concentrated in the first half of the Brillouin
zone between the {\it holon} and the {\it spinon} band. 
Although the relative intensity of the ARPES signal
at the two singularities depends on the details of the band structure, the 
power-law nature of the divergences implies that the intrinsic width of each peak 
is always comparable with the separation between the holon and the spinon branch
$\Delta\omega\sim \epsilon_h(k+\frac{\pi}{2})-[\epsilon_h(\pi)+\epsilon_s(\pi+k)]$. The
average intensity can be estimated on the basis of the sum rule (\ref{completeness}) and 
scales as $(\Delta\omega)^{-1/2}$, getting smaller when the two branches separate, as 
shown both in experiments \cite{kim06} and in numerical calculations \cite{Matsueda}.

\section{Conclusions}
\label{conclusions_sec}
The single spinon approximation, combined to Bethe Ansatz results and Lanczos diagonalizations
allows to obtain very accurate results for the dynamic properties of a single hole
in the one dimensional Hubbard model. The Lehmann representation of the spectral function (\ref{ako})
shows that two separate ingredients combine to define the overall shape of $A(k,\omega)$:
the excitation spectrum and the quasi-particle weight. The idea at the basis of our
method is to limit the size effects that plague numerical results by dealing with
these two quantities separately: in the 1D Hubbard model the excitation spectrum 
is given exactly by the Bethe Ansatz equations in the thermodynamic limit \cite{LW}, while 
the quasi-particle weight is obtained, in the single spinon approximation, by
Lanczos diagonalization. Size effects are shown to be negligible and the accuracy 
of the approximation can be checked {\it a posteriori} by a frequency sum rule (\ref{completeness}).  
Our expression for the spectral function of the 1D Hubbard model (\ref{a3}) is consistent 
with the structure predicted by bosonization \cite{ET} at weak coupling, provided the quasi-particle weight 
${\mc Z}_k(Q)$ factorizes as shown in Eq. (\ref{zfact}). A numerical test carried out at $U=3t$ 
does not show a convincing quantitative agreement with the result obtained by the form factor
approach \cite{ET}, possibly due to the difficulty to achieve the $U\to 0$ limit. 

The extension to finite doping is straightforward but 
in principle this approach can be also generalized to other fermionic lattice models, in one
or more dimensions, provided the relevant states entering the quasi-particle weight in
the strong coupling limit can be easily classified. This would be the case
of the extended Hubbard model (i.e. a Hubbard model with nearest neighbor Coulomb repulsion) or 
in the presence of lattice dimerization. 
Clearly, for non integrable models, no analytical information 
on the excitation spectrum is available and a size scaling on the energy spectrum is 
also required. A study of such generalizations may be useful to understand the
role of some perturbation on the spectral function of correlated electron models. 

The specific example of SrCUO$_2$ shows that 
our method allows for a direct comparison between theory and ARPES experiments and
for an accurate determination of the Hubbard parameters which best describe the 
hole dynamics in the material. The spectral function derived here
provides a natural explanation of the observed reduction of the spectral weight in a half
of the Brillouin zone and of the broad line-shape detected in experiments.
Future applications of this method to the case of cold atoms in optical traps may 
help in pointing out the peculiar features of one dimensional physics in other
experimental realizations of correlated one dimensional Fermi gases. 

We thank C. Kim and F.H.L Essler for stimulating correspondence.


\begin{thebibliography}{99}
\bibitem{noteU} The limitation to intermediate and strong coupling is a technical one:
at weak coupling size effects become more relevant and  the adiabatic procedure 
employed in this work might fail.
\bibitem{note}In one dimension and exactly at $U=\infty$ all spin configurations
are degenerate: however in the $U\to \infty$ limit the spin degeneracy
is lifted and the ground state of the Heisenberg model is singled out.
\bibitem{noteQ} Our definition of the spinon momentum $Q$, which for $U\to\infty$ must 
reduce to the total momentum of the Heisenberg wave function in Eq. (\ref{s}), differs from
the one usually adopted in Bethe Ansatz studies ($p_s$) \cite{woy}: $Q=\frac{3}{2}\pi - p_s$.
\bibitem{note3} In experiments the lower branch is not detected because it usually overlaps with 
other valence bands of the compound.
\bibitem{notek} The accuracy of the single spinon
approximation is similar also at weaker coupling: for $U=4t$ 
the completeness condition (\ref{completeness}) is violated at most by $0.015$ at all $k$'s.
\end{thebibliography}
\end{document}